\documentclass[aps,prl,preprint,groupedaddress]{revtex4-1}
\usepackage{graphicx}
\usepackage{mathptmx}
\usepackage{amsmath}

\begin{document}

\title{\textbf{\textit{Metashooting}: A Novel Tool for Free Energy Reconstruction from Polymorphic Phase Transition Mechanisms}}

\author{Samuel Alexander Jobbins,\textit{$^{a}$} Salah Eddine Boulfelfel,\textit{$^{b}$} and Stefano Leoni$^{\ast}$\textit{$^{a}$}}
\affiliation{$^{a}$~School of Chemistry, Cardiff University, Cardiff, CF10 3AT, Wales, UK\\
$^{b}$~Georgia Institute of Technology, School of Chemical and Biomolecular Engineering, Atlanta, GA 30332-0100, USA}

\begin{abstract}

\bf{We introduce a novel scheme for the mechanistic investigation of solid-solid phase transitions, which we dub \textit{metashooting}. Combining transition path sampling molecular dynamics and metadynamics, this scheme allows for both a complete mechanistic analysis and a detailed mapping of the free energy surface. This is illustrated by performing \textit{metashooting} calculations on the pressure-induced B4/B3 $\rightarrow$ B1 phase transition in ZnO. The resulting free energy map helps to clarify the role of intermediate configurations along this activated process and the competition between different mechanistic regimes with superior accuracy. We argue that \textit{metashooting} can be efficiently applied to a broader class of activated processes.}

\end{abstract}

\maketitle

\section*{Introduction}
Molecular dynamics (MD) has become the method of choice to investigate complex chemical phenomena at the atomistic level. Thanks to both efficiently written algorithms and high-performance computational infrastructures, the scope of numerical simulations has markedly widened. Nonetheless, the presence of inherent kinetic bottle-necks means that entire regions of configuration space may remain unexplored. If such states are separated by large energy barriers, even the most efficient MD will end up spending considerable time within long-lived, metastable states. Furthermore, depending on the height of those barriers, the event of interest may fully elude observation over the time scale of the simulation~\cite{Chandler:1978hh, Dellago:2006bg, Laio:2002ft}.\\

Protein folding, nucleation, crystallisation, phase transitions and chemical reactions are all important phenomena which are characterised by activated steps. Their detailed mechanistic investigation would benefit considerably from a precise mapping of the underlying free energy landscape. However, despite improving numerical algorithms and increasing computational power, their systematic investigation remains extremely demanding and often completely impractical. Methods to explore multi-minima energy landscapes exist~\cite{Glover:1990ks,Sibani:2007fc} and recently there has been a pronounced effort to develop novel approaches to overcome this time-scale problem~\cite{Laio:2002ft, Dellago:1998kw}. Historically, the use of an external bias aimed at enhancing the frequency of rare events in MD simulations is realised in methods such as Umbrella Sampling (US)~\cite{TORRIE:1977hs}. This is achieved by systematically adding a bias along well-chosen coordinates, in order to sample only small but overlapping regions from which a free energy can be reconstructed. Another bias-oriented method, metadynamics (metaD)~\cite{Laio:2002ft} exploits the concept of Collective Variables (CV), which span a lower dimensional coordinate space. Within the space of these CVs, a history-dependent bias is deposited, which rapidly drives the system away from its initial, probable state. The definition of a good set of CVs is therefore key to the success of the method~\cite{Abrams:2014gh}. Within a CV space, standard MD would linger around a minimum close to the initial state corresponding to some value of the CV. Overcoming free-energy barriers by metadynamics inevitably leads to broader exploration of CV values by crossing over to different states within a finite computational time.

Metadynamics takes place within the space of one or a set of CVs, which have to be known at the beginning of the calculation. A valid set of collective variables optimally covers the entire space of conformations of interest. This clearly requires that CVs be engineered with the global free energy map in mind. However, this is evidently problematic, as often the underlying free energy landscape is the desired final result of the simulation and can often not be deduced by intuition. Additionally, the requirement for such a `global' CV directly clashes with the effort to keep its dimensionality as low as possible.

Recently, a  novel approach has been introduced, which allows for a "bespoke bias" based on a variational approach~\cite{Valsson:2014jr, Valsson:2015ck}. Within this approach, the CV problem is partially overcome by reinterpreting its meaning as a local descriptor of metastable states. Enhancing static fluctuations will amplify the probability of a crossover into another metastable state. Importantly, the bias can be deposited in such a way that partial filling can be achieved in all basins of interest. While the success of the method still depends on an appropriate choice of CVs, this represents a novel general scheme for enhanced-sampling MD simulations.\\

To tackle the intrinsic disparity between the time scales of activated processes and molecular dynamics simulations, the Transition Path Sampling method (TPS)~\cite{Bolhuis:1998cj,Dellago:1998kw} was introduced. TPS is based upon the idea that, in a complex system, crossing a barrier between (meta-)stable states may happen in a multitude of different ways. In such a scenario, the potential energy surface of a system is unlikely to be represented by saddle points alone. Instead, numerous points on the potential energy surface become relevant, some of which may still be stationary points, but many others not. The implementation of the method consists of collecting such an ensemble of reactive trajectories, without previous knowledge of the details of the process (that is, no \textit{a priori} information or guess about the reaction coordinates is required - such information should naturally result from the simulation approach). Within a transition ensemble, the relevance of a path is weighted by its probability. TPS collects true dynamical trajectories and can be plugged on top of different molecular dynamics schemes, within different simulation ensembles~\cite{Dellago:2006bg}.

TPS belongs to the class of pathway-orientated methods and allows for an exquisite level of detail in mechanistic investigations~\cite{Boulfelfel:2008fx,Leoni:2004eb,Zahn:2004he,Boulfelfel:2007do,Leoni:bm,Boulfelfel:2012iq,Boulfelfel:2012kr}. Different from metadynamics, TPS does not deposit any bias potential; similarly to metadynamics, it deploys a low-dimensional order parameter, which distinguishes between the basins of interest. TPS is designed to enhance statistics in intermediate regions between long-lived basins, however it is generally not a method of choice for free energy reconstruction as the main statistical weights are still contributed by (meta-)stable basins.

Here, we present a novel combination of TPS and metadynamics techniques, which we dub \textit{metashooting}. This novel approach allows for reconstruction of the underlying free energy of a process from TPS trajectories. This combination is based on the propagation of CVs \textit{from TPS into metaD}. While TPS controls the rapidity of CV variation (inherent to true dynamical trajectories), metadynamics deposits a bias around regions of the CV that vary comparatively slowly. In doing so, TPS keeps the path crossing probability ratios (A $\rightarrow$ B to B $\rightarrow$ A) unchanged, thus preventing unilateral over-biasing. By rapidly varying the controlling CVs in TPS, partial basin filling can be achieved, while maintaining the intermediate regions bias-free for most of the calculations. Only in its final phases do the details of the intermediate regions appear.

\begin{figure}[t]
 \centering
 \includegraphics[width=9cm]{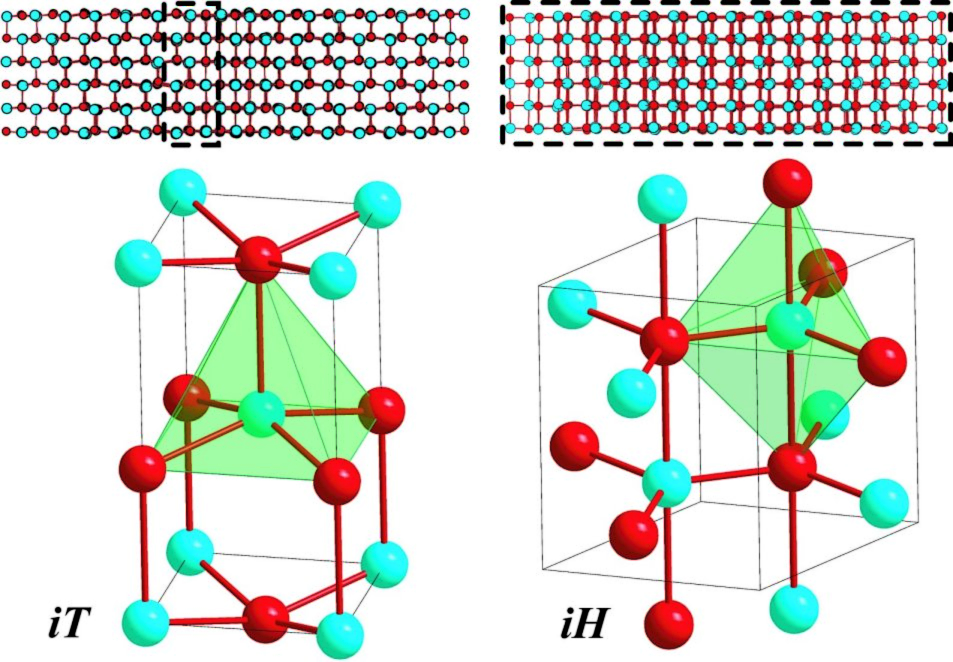}
 \caption{Two 5-coordinate intermediates (tetragonal \textit{iT} (\textit{I4mm}, left) and hexagonal \textit{iH} (\textit{P63/mmc}, right)) were shown to compete against one another in the B4-B1 transformation. Whilst the \textit{iT} was only found at the interface between B4 and B1, the \textit{iH} structure sometimes dominated the entire system under study. Despite this, \textit{iH} phase was shown not to be crucial to the transformation~\cite{Boulfelfel:2008fx}.}
 \label{Fig1}
\end{figure}

In the following we illustrate the details of our approach by performing a complete \textit{metashooting} analysis of the solid-solid reconstructive phase transition between the B3  zincblende phase and the B1 high-pressure rocksalt phase of zinc oxide (ZnO). ZnO has proven to be extremely challenging over the years~\cite{Zagorac:2012jy,Catlow:2008ke,Saitta:2004cm,Bayarjargal:2014bk,Cai:2007hr}, with several attempts to conclusively shed light onto competitive mechanisms and intermediates. Saitta and Decremps first discussed the competitive nature of the `tetragonal' and `hexagonal' routes in the B4-B1 phase transition, and their work concluded that semiconductors involving $d$-electrons, such as ZnO, preferred the `tetragonal' path~\cite{Saitta:2004cm}. However, other work seemed to contradict this - notably the experimental work of Liu \textit{et al.}, which utilised high resolution angular dispersive X-ray diffraction, which seemed to indicate that the `hexagonal' pathway was the more favoured for ZnO at lower pressures~\cite{Liu:2005hb}. Second harmonic generation experiments pinpointed the role of the pressure medium (hydrostatic vs. non-hydrostatic) to promote the formation of centrosymmetric $iH$ or acentric tetragonal $iT$, respectively~\cite{Bayarjargal:2014bk}. In a previous work~\cite{Boulfelfel:2007do}, TPS was used to produce a detailed mechanistic pathway for the B4-B1 transition, and it was shown that the mechanism proceeds via a rich series of transition states and intermediates~\cite{Boulfelfel:2008fx} Additionally, the two 5-coordinate intermediate motifs were visited and were shown to be competing with one another in the mechanism. The analysis, based on dynamical trajectories, concluded that a hexagonal intermediate could indeed be visited, however it was not essential to the transformation. In fact, the reconstruction was found to propagate over interfaces of local tetragonal motifs (Fig.\ref{Fig1}). 

It is therefore apparent that the phase transitions between different polymorphs of ZnO are very non-trivial. Indeed, much debate still rages over their precise nature and work is still being published to attempt to unravel the minutia of these elusive mechanisms, with a particular emphasis on the B4-B1 transformation. 
In the following, after a summary of implementation details, the results of \textit{metashooting} on a challenging solid-solid phase transition in this material between the B3 and B1 phases are discussed.
\\
\section*{Methods}
Our complete analysis of the B3-B1 phase transition in ZnO proceeds over 2 steps: i) TPS calculations, including the definition of a suitable order parameter and ii) free-energy mapping with \textit{metashooting}, where TPS and metaD are combined.

\subsection*{Transition Path Sampling}

The aim of TPS is to obtain the transition path ensemble~\cite{Bolhuis:1998cj,Dellago:1998kw} and an efficient approach to this task involves the use of Monte Carlo techniques. A random walk is carried out in the space of the trajectories, whereby a new path $x^{(n)}(\tau)$ is generated from an old one, $x^{(o)}(\tau)$. The implementation of the random walk moves are based on the shooting algorithm~\cite{Dellago:2006bg}, whose general premise is to generate pathways by "shooting off" a new pathway from a randomly selected time slice $x^{(o)}_{t^{\prime}}$ of the previous pathway. Modifications are introduced to $x^{(o)}_{t^{\prime}}$, yielding $x^{(n)}_{t^{\prime}}$. From this perturbed time slice at time $t^{\prime}$, two new path segments are generated forward to time $\tau$ and backward to time $0$, in molecular dynamics runs within the chosen ensemble. A trajectory is rejected if the introduced modifications cause the new trajectory not to connect the basins of interest. Otherwise, it is accepted with a given probability~\cite{Dellago:2006bg}. The acceptance/rejection criteria thus depend on the shooting point only, and can be easily translated into an algorithm. The introduction of such a probability weighting scheme allows the system to cross intermediate activation barriers of the order of $k_{B}T$ or lower. Perturbations are introduced as momenta modifications under strict conservation of total linear and angular momenta.

\subsubsection*{Order Parameter}

Transition path sampling calculations require the initial and final configurations of the system to be strictly distinguishable, such that there is no overlap of their respective basins when projecting onto the order parameter~\cite{Bolhuis:2002ew}. A handy yet powerful order parameter used in previous path sampling calculations is the average coordination number of the system. The coordination sphere (CS) is a set of integers ${n_1,n_2,\dots n_i}$, in which the \textit{i}$^{th}$ term is the number of atoms in "shell" $i$ that are bonded to atoms in "shell" $i-1$. Shell 0 consists of a single atom, and the number of atoms in the 1$^{st}$ shell $n_1$ is the conventional coordination number (CN). This quantity can be calculated for all atoms in a system, collated, and averaged over the whole structure. The 1$^{st}$ three shells of the coordination spheres of the three phases of zinc oxide are: zincblende (B3) ${[4,12,24]}$, wurtzite (B4) ${[4,12,25]}$ and rocksalt (B1) ${[6,18,38]}$. The 1$^{st}$ coordination number (1$^{st}$ CN) of zincblende and rocksalt are 4 and 6, respectively. During TPS, the coordination spheres of each atom were calculated up to the 3$^{rd}$ shell, but only the 1$^{st}$ CN was utilised to trace the progress of the path sampling calculations. The 2$^{nd}$ and 3$^{rd}$ CNs were recorded simply to gain further insight into the underlying details of the transition pathways.

Using the 1$^{st}$ coordination shell only means that there is no discrimination between the B3 and B4 configurations. As the system is starting from a metastable B3 structure (see below), changes in the final regime can be expected. Furthermore, using solely the 1$^{st}$ coordination number as the order parameter exerts absolutely no bias on the calculation. The variation in the order parameter arises simply as a result of the application of the transition pressure to the system. Hence, monitoring the 1$^{st}$ coordination sphere does not force the system to evolve in any particular way over the course of the path sampling iterations. In TPS, both the overall mechanism and the trajectory endpoints are rectified~\cite{Leoni:2008ff}. Therefore, the order parameter is chosen with minimal constraints in mind, in order to allow the process to express the most probable outcome. A higher-order CS would drive the system into a specific transformation regime within a \textit{biased TPS} approach, in order to study specific, less probable transformation pathways.   

\subsection*{\textit{Metashooting}}

\textit{Metashooting} combines TPS and metadynamics to efficiently reconstruct the underlying free energy landscape from TPS trajectories. This way, computing time will not be expended by filling other regions of configuration space that had no relevance to the transition.

The general procedure of the \textit{metashooting} approach for a transition A $\rightarrow$ B is as follows:

\begin{enumerate}
\item \textbf{TPS step 1}: Starting from a randomly chosen snapshot along the TPS trajectories, shoot off a new trajectory and time propagate.

\item \textbf{MetaD step 1}: Once basin A is reached, deposit a bias using a metadynamics scheme (in this case, using the well-tempered approach), in order to locally partially fill the as-yet uncharacterised energy well. At this stage, a small number of metasteps are chosen, which allows the CV to vary about the characteristic value of the basin.

\item \textbf{TPS step 2}: Generate another trajectory (as in step 1) and propagate into the now biased basin A, to generate a true MD velocity distribution on the biased potential.

\item \textbf{TPS step 3}: Similarly to TPS step 1, shoot off a new trajectory from the same snapshot as above and propagate, only this time backwards in time, to basin B.

\item \textbf{MetaD step 2}: In basin B, apply the same metadynamics scheme as above for a number of steps, in order to partially fill this second basin of attraction.

\item \textbf{TPS step 4}: Re-shoot a novel trajectory from a random snapshot of the previous trajectory. Propagate into the now biased basin B.

Repeat the above iterations of metadynamics and TPS moves, until the underlying free energy profile is fully converged.

\end{enumerate}

The number of metadynamics steps for each \textit{metashooting} move can be chosen such that only moderate variations around the current CV value are allowed, as a ${\Delta CV}_{Max}$. Alternatively, as a better means to control bias deposition without imbalance, TPS path acceptance probability in TPS can be chosen to stay the same, i.e. at no point can crossing over the activation energy barrier become easier for either direction $A \rightarrow B$ or $B \rightarrow A$. This is realised within an adaptive scheme of rescaling momenta modifications (see below).

\subsection*{Calculation details}

\subsubsection*{Molecular Dynamics}
Born-Oppenheimer molecular dynamics simulations were performed using the \textit{cp2k} package within the $NpT$ ensemble at $T$ = 300 K and $p$= 9.8 GPa. At this pressure, the enthalpy of B1 and B3 are equal. Constant pressure and temperature were ensured by the Martyna-Tobias-Klein~\cite{Martyna:1992gy} algorithm, allowing for anisotropic shape changes of the simulation box~\cite{Martyna:1996ga}. Inter-atomic forces were calculated from a Buckingham pair potential~\cite{BINKS:1993dv}, which has been previously validated~\cite{Boulfelfel:2007do}. Long range electrostatic effects were accounted for using an Ewald summation. Time propagation was performed with a time-reversible predictor-corrector algorithm. An integration timestep of 0.2 fs was used. The simulation box contained 1200 Zn-O atom pairs.

\subsubsection*{Transition Path Sampling}

The sampling starts from an initial trajectory connecting the limiting phases. A new trajectory is generated by selecting a configuration from the existing one and slightly modifying the atomic momenta. The modifications $\delta$p are applied to randomly chosen pairs of atoms $(i,j)$ according to: $\vec{p}_{new} = \vec{p}_{old} + \delta p \cdot (\vec{r}_{j}-\vec{r}_{i})/ \lvert \vec{r}_{j}-\vec{r}_{i} \rvert$ and $\vec{p}_{new} = \vec{p}_{old} - \delta p \cdot (\vec{r}_{j}-\vec{r}_{i})/\lvert \vec{r}_{j}-\vec{r}_{i} \rvert$, keeping both momentum and angular momentum conserved. The resulting atomic momenta are rescaled by a factor of $\sqrt{E_{kin}^{old}/(\sum_{i=1}\lvert \vec{p}_{i} \rvert ^2 / 2m_i)}$, in order to keep the total kinetic energy conserved. The propagation of the modified configuration in both directions of time $(-t,+t)$ generates a new trajectory. Repeating this step provides a set of trajectories, with successful examples representing the transition ensemble. TPS requires an initial guess trajectory, which we obtained from a topological/geometric approach of interpolating between the two limiting crystal structure (B1 and B3), described elsewhere~\cite{Leoni:2004eb, Zahn:2004he, Leoni:2008ff}. Momenta modifications were introduced using an adaptive scheme, which entailed rescaling $\delta$p to larger or smaller values, based on whether this $\delta$p led to a successful or failed trajectory, respectively. The acceptance ratio for the production runs was generally between 40-60\%.  

\subsubsection*{Metadynamics}

Metadynamics calculations were carried out using the \textit{plumed} plug-in~\cite{Tribello:2014eb}, in which a history-dependent Gaussian bias is added to the underlying potential. The well-tempered~\cite{Barducci:2008ji} prescription of metadynamics was used, in which the height of the Gaussian hills are rescaled at each step. The collective variables corresponded to the 1$^{st}$ and 3$^{rd}$ coordination spheres of the zinc oxide systems - analogous to the order parameter used in TPS. The coordination sphere in plumed is mapped onto a switching, differentiable function, $s(r)=\frac{1-(\frac{r-d_0}{r_0})^n}{1-(\frac{r-d_0}{r_0})^m}$. $r$ is a Zn-O inter-atomic distance, while $r_0, d_0, m, n$ are adjustable parameters. For the 1$^{st}$ and 3$^{rd}$ coordination sphere the parameters were set to (2.6, 0.1, 6, 12) and (5.3, 0.1, 6, 12), respectively. These parameters realise a one-to-one correspondence between coordination numbers calculated geometrically and values based on the switching function, despite resulting in slightly exaggerated values for the 3$^{rd}$ coordination sphere of B4/B3 (28/29 instead of 24/25). Gaussian functions were deposited every 500 simulation steps. On the average, 5 metasteps were calculated on each side of the trajectory during each iteration of the \textit{metashooting} procedure. The height of the Gaussians was set at 1000 kJ mol$^{-1}$. The Gaussian widths were chosen in accordance with the variance of the CV in an unbiased simulation - as such, these were set to 0.2 and 1.0 for the 1$^{st}$ and 3$^{rd}$ coordination number respectively. The well-tempered `bias-factor' was set at 10000, which is large enough to escape minima and re-scale deposited Gaussians within a meaningful time frame.

\section*{Results}

\subsection*{Mechanistic analysis}

TPS was initiated from a B3-B1 trajectory regime, obtained from the modelling approach described and used elsewhere~\cite{Leoni:2000vw,Leoni:2004eb}. One of the advantages of initiating the process from a plausibly designed trajectory is that it takes fewer TPS iterations for the system to move away from the initial, unfavourable regime to a more probable one. The process of moving towards a probable regime is known as \textit{trajectory decorrelation}. In fewer than 50 TPS iterations, the procedure steered the trajectory away from the collective motion encoded by the geometric model towards a regime characterised by nucleation events. The trajectories continued to evolve until no further major mechanistic changes were observed, which occurred after several hundred iterations of the path sampling procedure.\\

\begin{figure}[t]
 \centering
 \includegraphics[width=9cm]{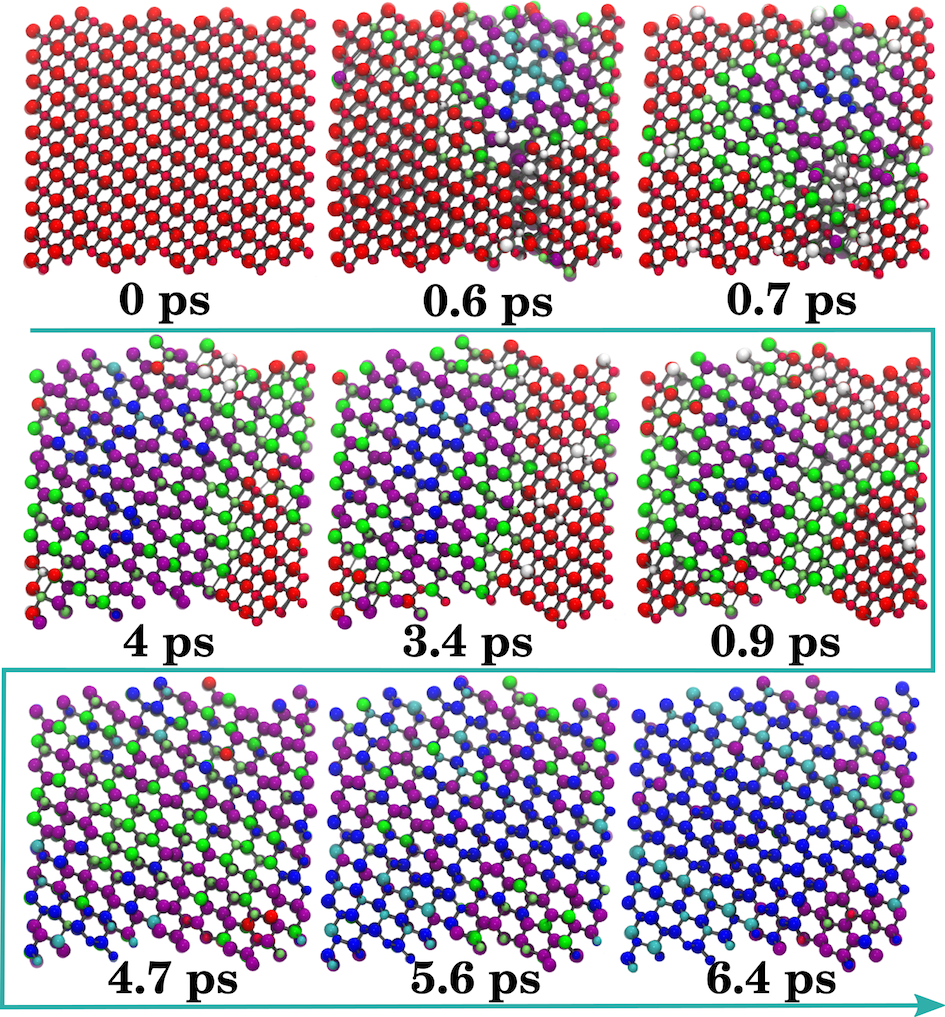}
 \caption{Coordination-coloured snapshots from a representative trajectory, showing a sequence from rocksalt (B1) to a mixed wurtzite-zincblende (B4/B3) configuration. A nucleus forms in B1 within 500 fs, followed by shearing along (111)$_{B1}$ at 0.6-0.7ps. The nucleus grows further over intermediates characterised by 5- and 4-fold coordination (0.9-4.7 ps), until the system transforms into a 4-coordinate motif, which is composed primarily of hexagonal wurtzite (5.6-6.4 ps). Colour Code: red: B1, green: \textit{iH, iT}, dark blue: B4 (wurtzite), light blue: B3 (zincblende), purple: 4-coordinated, but not corresponding to B3 or B4}
 \label{Fig2}
\end{figure}

Fascinatingly, the final 4-coordinated product was not a pure zincblende (B3) structure, but a mixed B4/B3 system dominated by the wurtzite (B4) form. In all successful trajectories, the final material consists of only 10-20\% B3, with the remainder of the structure formed by the more stable hexagonal form. The order parameter based on the 1$^{st}$ coordination number permits both the wurtzite and zincblende configurations. Whilst the initial, unfavorable regime was B1 $\rightarrow$ B3, TPS rectified the mechanism into B1 $\rightarrow$ B4/B3, which contained predominantly hexagonal components.

The complexity of the system leads to several types of trajectories, with some degrees of (dis)similarity in the details of the reconstruction. 

A representative B1 $\rightarrow$ B4/B3 trajectory is shown in Fig.~\ref{Fig2}. A nucleus of 4-connected ZnO (Fig.~\ref{Fig2}, 0.6 ps) forms within B1, followed by the formation of regions of 5-fold coordination at the interface with B1 (Fig.~\ref{Fig2}, 0.7 ps), which are subsequently outgrown by 4-fold coordinated regions (Fig.~\ref{Fig2}, 0.9-4.7 ps). Regions of 5-fold coordination (mostly \textit{iH}) are found between the expanding 4-connected region and B1. The transformation into a mixed B4/B3 entails complete resorption of areas of 5-fold coordination, Fig.~\ref{Fig2}, 4.7-6.4 ps. The transformation of remaining B1 motifs (4 ps) is associated with a sharp volume expansion, which causes the box volume to fluctuate, under appearance of local 5-fold motifs, Fig.~\ref{Fig2} 4.0-4.7 ps.

From a configuration of the type seen in Fig.~\ref{Fig2} at 0.6 ps, a less severe compression leads to the growth of a 5-fold region at the expense of any 4-connected one. This accompanies the formation of a mostly 5-coordinated intermediate containing residual B1, denoted B1-\textit{iH}. This configuration persists for as long as 500 ps in subsequent molecular dynamics simulations. The presence of an intermediate, which is not vital to the transformation and corresponds to a 5-membered intermediate along the transition pathway, was noticed in previous work~\cite{Boulfelfel:2008fx}. The nature of this intermediate is the hexagonal \textit{iH} form, whilst the tetragonal analogue \textit{iT} only appears as a motif at interfaces during the reconstruction.

\begin{figure}[t]
 \centering
 \includegraphics[width=9cm]{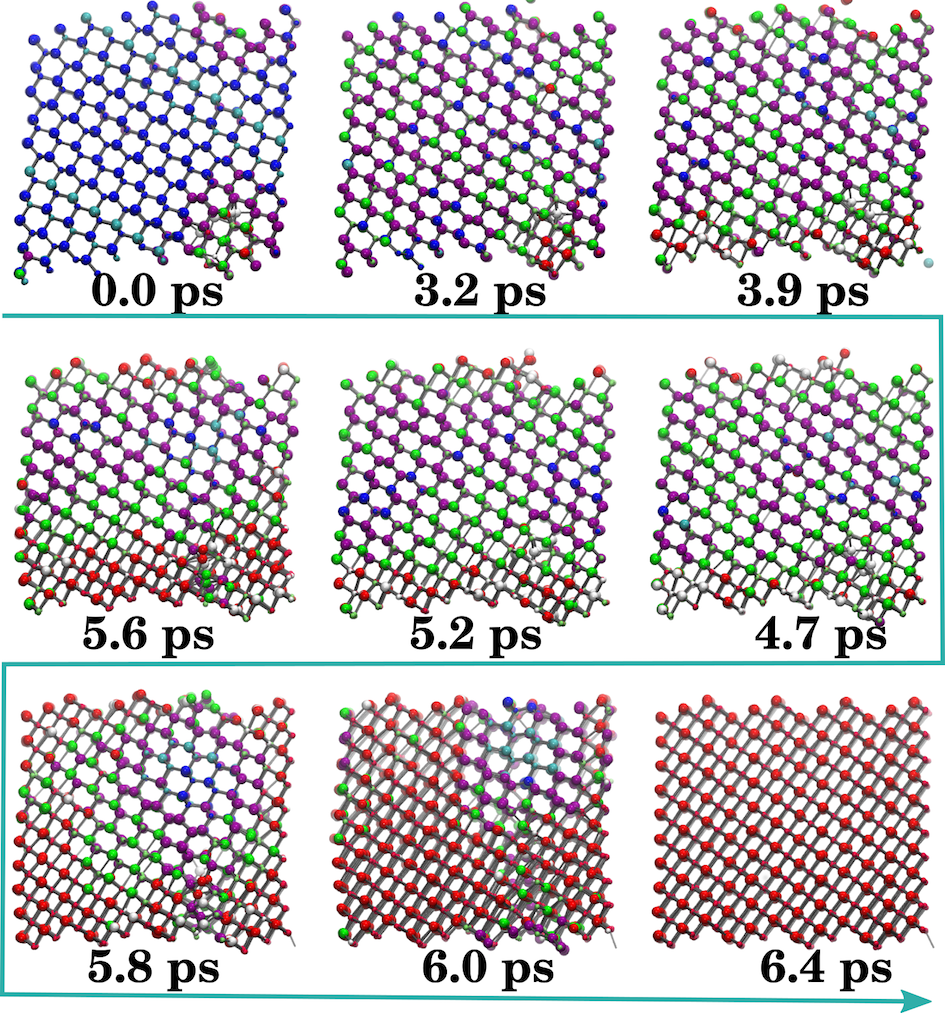}
 \caption{Coordination-coloured snapshots of a representative B4/B3 $\rightarrow$ B1 trajectory. Mixed (B4/B3) wurtzite-zincblende (0.0 ps) is followed by a 4-coordinated distorted configuration (0.0-3.2 ps), purple atoms. Subsequent growth of 5-coordinated areas (green), mainly of type \textit{iH} are observed (3.2-5.2 ps), from which B1 (red) develops (5.2-6.0 ps) until the structure locks-in into 6-coordinate B1 (6.4 ps). Colours as in Fig.~\ref{Fig2}}
 \label{Fig3}
 \end{figure}

The B4/B3 $\rightarrow$ B1 mechanism slightly deviates from the reverse process (Fig.~\ref{Fig3}). In this trajectory type, the starting 4-coordinate material is first collectively distorted (Fig.~\ref{Fig3}, 0.0-3.2 ps, purple atoms). From 5-coordinated areas of \textit{iH} motifs (Fig~\ref{Fig3}, 3.9-5.2 ps), the B1 structure develops (Fig.\ref{Fig3}, 5.2-6.4 ps). Thus, over half of the transformation time is spent slowly squeezing the system until it can transform into 5-coordinated \textit{iH} regions, indicating the presence of a deep but fairly gently-sloping energy barrier separating the B4/B3 basin from the intermediate configurations. In all configurations, \textit{iH} motifs are relatively long-lived. This is not the same as the reverse B1 $\rightarrow$ B4/B3 transition, in which \textit{iH} regions were less extensive and promptly eliminated at interfaces (Fig.~\ref{Fig2}), with the majority of the simulation time being dominated by coexisting 4- and 6- coordinated motifs. 

Consistent with previous calculations~\cite{Boulfelfel:2008fx}, different intermediates are associated with different types of trajectories. The presence of B3 adds further structures to an already complex scenario. In previous work, low transition pressure (less than 10 GPa) were noticed to favour the appearance of $\textit{iH}$, which benefits from a less severe [0001]$_{B4}$ compression. The results presented here appear to corroborate that view.

\subsection*{\textit{Metashooting}}

In order to better understand the competitive nature of different mechanisms, the free energy (FE) was evaluated by coarse-graining the potential energy surface onto a suitable set of collective variables. Free energy surfaces have found extensive use for the description of biological systems (in particular peptide calculations~\cite{Granata:2015jm}), but are less frequently used in the solid state.

In \textit{metashooting}, mechanistic details acquired from the transition path sampling procedure remain unspoiled. As the process is driven by TPS trajectories, metastable basins are filled in a balanced way - starting from the two limiting configurations, and finishing with the characterisation of all relevant intermediate configuration. Repeated iterations of the \textit{metashooting} algorithm (see Methods) enables the free energy landscape corresponding only to the courses of the TPS-derived trajectories to be built up in a step-wise and equally distributed manner. As a result, regions around the transition states are considered only at the end of the analysis, avoiding any potential bias that may scramble the mechanistic details.

The algorithm should first fill up the two starting basins of attraction from the final TPS trajectories (B4/B3 and B1). Visiting these regions should become discouraged as a result of the bias potential. The progress of \textit{metashooting} can be monitored by visualising the Gaussian deposition. Eventually, the \textit{metashooting} algorithm should fully characterise the central (around metastable basins) and peripheral regions of the CV space over the course of the trajectory. Convergence will be seen in a similar way to a standard metadynamics scheme - that is, the underlying free energy landscape should no longer significantly change with further iterations of the procedure. At this point, meaningful analysis of intermediates and energy barriers associated with each process can be carried out.\\

Harnessing the free energy in this way should have a number of advantages. The landscape can be thought of as the surface upon which trajectories travel, and individual characteristics of the trajectories are determined by which regions of the surface are visited. Thus, its elucidation allows for a complete quantitative analysis of the underlying thermodynamics and kinetics of the process; measurement of the activation energy and change in free energy between different steps of the procedure, as well as the starting and finishing products, can be trivially taken from the plot. Moreover, intermediates and transition states that are vitally important to the transformation will appear obvious along the reaction scheme. Such insight into trajectories using path-based analyses of transformations only is not trivial. 

The free energy profile was constructed from converged TPS trajectories. After 250 iterations of the \textit{metashooting} procedure, the free energy had fully converged into a detailed landscape.\\

\subsubsection*{Progression of the \textit{metashooting}}
The evolution of the free energy landscape according to the \textit{metashooting} algorithm is first discussed, followed by an analysis of the final profile generated by the procedure.

\begin{figure}[t]
 \centering
 \includegraphics[width=9.2 cm]{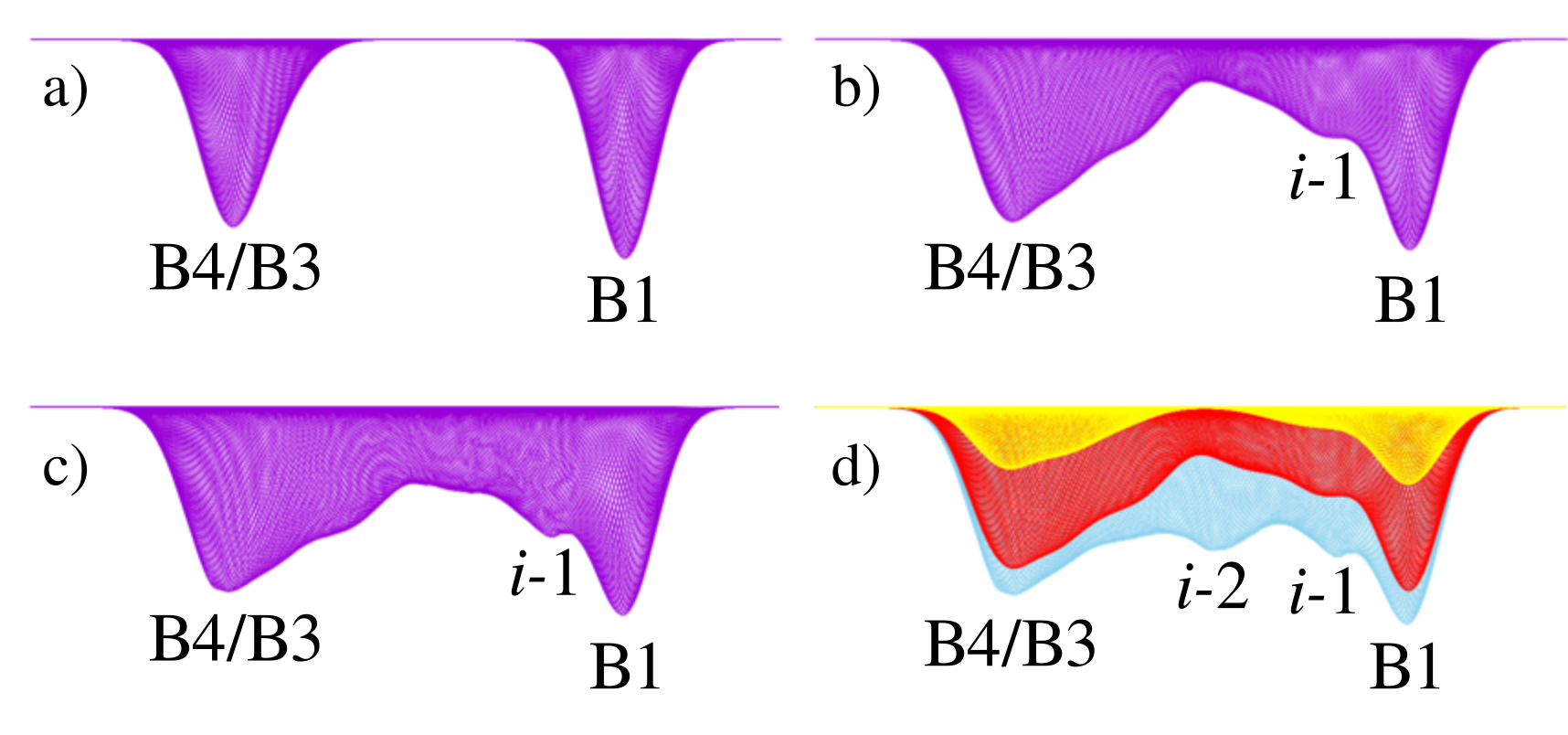}
 \caption{Free energy landscapes at different stages of \textit{metashooting}. a) Filling of the limiting basins centred on B4/B3 and B1 motifs (1$^{st}$ \textit{metashooting} cycle), b) appearance of a minimum \textit{i-}1 after 50 cycles, c) intermediate areas are explored after 200 cycles. In d) stages 10 (yellow), 75 (red) and 225 (blue) are displayed. In the latest cycles only the intermediate \textit{i-}2 appears, which corresponds to B1-\textit{iH}. The 3D free energy is plotted as a function of the 1$^{st}$ and 3$^{rd}$ coordination number. See text for further details.}
 \label{Fig4}
\end{figure}

The first few iterations of the procedure (Fig.~\ref{Fig4}) accumulated Gaussian functions only within the two limiting basins (B4/B3 and B1). A small `shoulder-peak' formed in the 4-coordinate basin, as a result of the oscillation between different forms of the low coordination material, including B3, B4 and other forms.\\

\begin{figure}[b]
 \centering
 \includegraphics[width=9.2 cm]{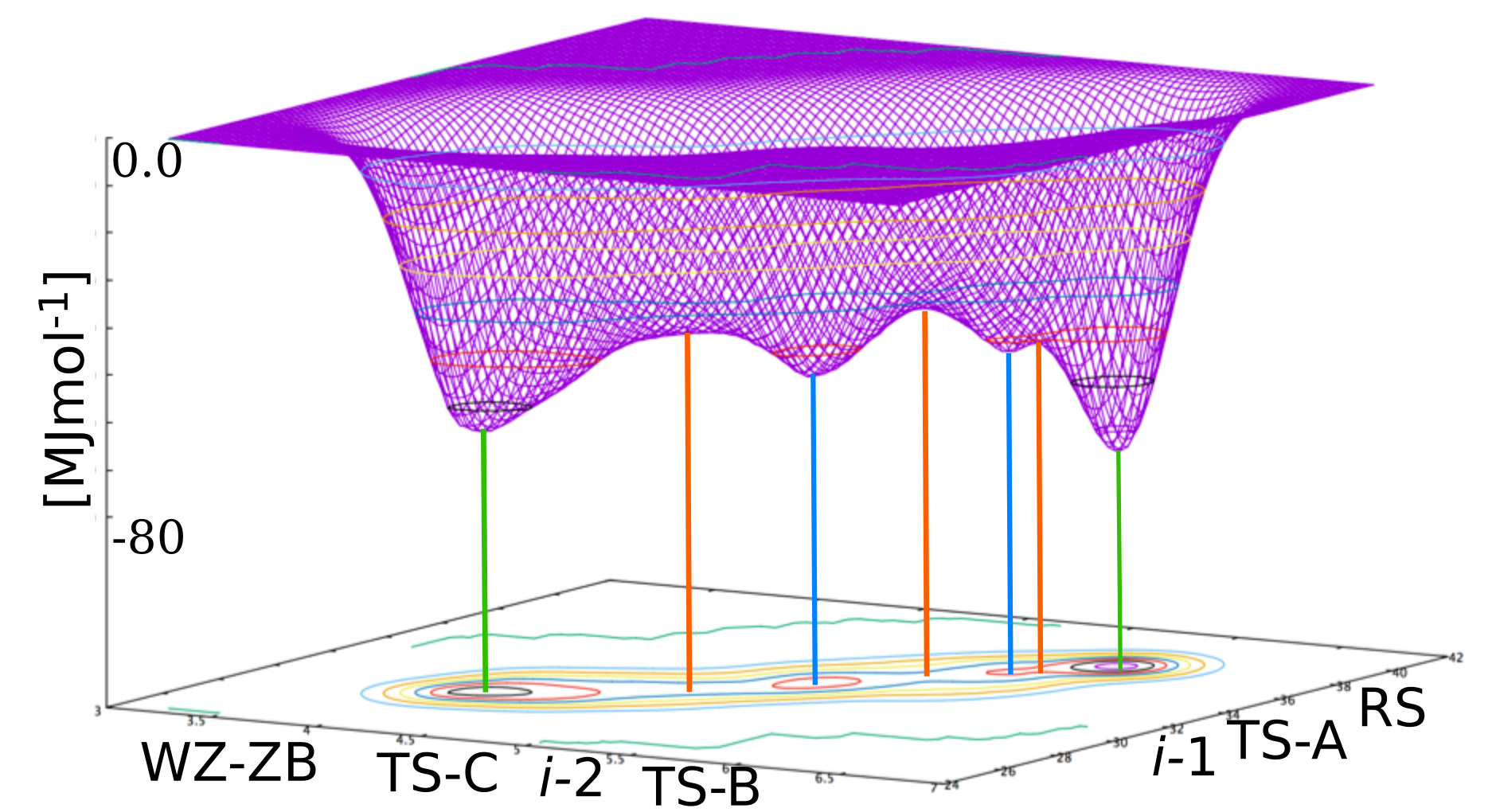}
 \caption{Three-dimensional plot of the converged free-energy surface for the transition after 250 iterations of the \textit{metashooting} procedure. Two main basins, WZ-ZB (B4/B3) and RS (B1), two intermediate basins (\textit{i-}1 and \textit{i-}2), as well as three transition states (TS-A, TS-B, TS-C) are visible.}
 \label{Fig5}
\end{figure}

After 50 iterations (Fig.~\ref{Fig4}b), a notable `shoulder-peak' was evident adjacent to the B1 basin. After 150 iterations, that had resolved into a \textit{bona fide} minimum \textit{i-}1. By iteration 200, a second minimum centred half-way between the two limiting basins started to appear (Fig.~\ref{Fig4}c), which had fully resolved by iteration 225 (Fig.~\ref{Fig4}d), denoted \textit{i-}2. By 250 iterations (Fig.~\ref{Fig5}), the plot was fully converged - the energy barriers and positions of minima remained unchanged with any further iterations of the \textit{metashooting} procedure. 
\\
\subsubsection*{Energy profile and intermediate configurations}

There are at least seven points of interest on the free energy surface - the two initial basins of attraction (corresponding to B4/B3 and B1), two intermediate basins and three maxima, corresponding to transition state regions. 
The structures and roles of these turning points in the transformation can be ascertained by obtaining the 1$^{st}$ and 3$^{rd}$ coordination numbers of the turning points on the graph and matching these up to structures along the hundreds of trajectories from both the `plain' TPS runs and the \textit{metashooting} trajectories. It stands to reason that these regions must play important roles in the transformation mechanism. 
\\
\subsubsection*{Free-energy intermediates and maxima}

\textit{i-}1 is the metastable basin adjacent to the B1 basin. Every trajectory congregates at \textit{i-}1 before proceeding along a single pathway to or from the rocksalt structure. Despite the small and shallow dimensions of the basin, it clearly plays a central role in the transformation. It is separated from B1 by TS-A. The configuration associated with this maximum is an almost entirely B1 structure at the moment of the deformation of the rocksalt motifs. For the first few Zn-O sites this entails overcoming an activation energy, as they leave their 6-fold coordination.

\begin{figure}[t]
 \centering
 \includegraphics[width=9.2 cm]{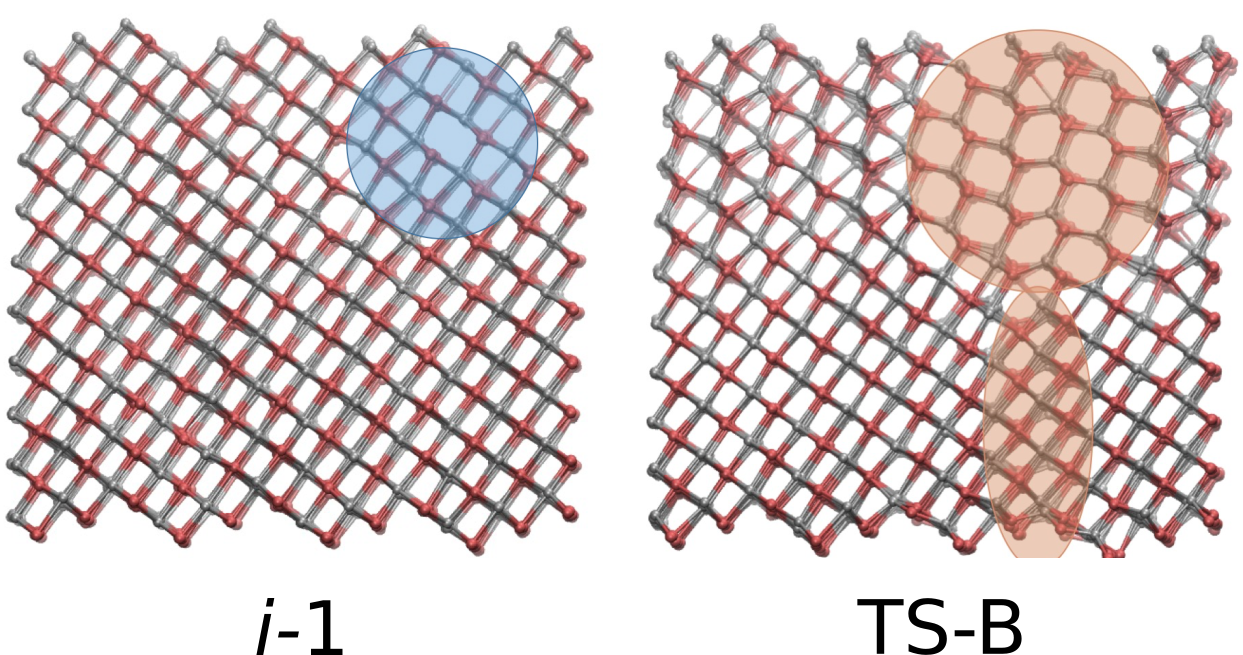}
 \caption{(left) Configuration corresponding to the small minimum \textit{i-}1, which is just after the very first rocksalt motifs locally deform into hexagons. (right) A representative configuration corresponding to the maximum TS-B, showing an enlarged seed accompanied by (111)$_{B1}$ layer shearing.}
\label{Fig6}
\end{figure}

In \textit{i-}1 a portion of the pristine B1 fragments contains 4- and 5-coordinated motifs. With respect to the way the square pattern is opened, the locally 5-coordinated motifs correspond to \textit{iH}. This small region is extremely short lived but present in every successful trajectory sampled, thus explaining its place on the free energy profile. Interestingly, this mechanistic pattern can not propagate as it is. The next free-energy barrier TS-B is overcome by including an additional (111)$_{B1}$ shearing mechanism, as seen in both the forward and reverse TPS trajectories. This configuration, which includes an enlarged seed region and the onset of the shearing, corresponds to the free-energy maximum TS-B. The configuration TS-B could easily be seen in the TPS trajectories, however the fact that this part of the transformation corresponds to two distinct events would have been completely missed without the \textit{metashooting} analysis. 
\begin{table}
\small
  \caption{\ 1$^{st}$ and 3$^{rd}$ coordination spheres (as defined by the switching function) at the centre of the regions defining the turning points.}
  \label{table1}
  \begin{tabular*}{0.5\textwidth}{@{\extracolsep{\fill}}lll}
    \hline
    Turning Point & 1$^{st}$ CN & 3$^{rd}$ CN \\
    \hline
    RS (B1) & 5.98 & 37.90 \\
    TS-A & 5.75 & 36.66 \\
    \textit{i-}1 & 5.68 & 36.21 \\
    TS-B & 5.43 & 34.86 \\
    \textit{i-}2 & 5.15 & 33.13 \\
    TS-C & 4.87 & 31.62 \\
    WZ-ZB (B4/B3) & 4.10 & 29.22 \\
    \hline
  \end{tabular*}
\end{table}

The second intermediate region, \textit{i-}2 is a deep energy well placed halfway between the two limiting basins. Trajectories visiting configurations characterised by 1$^{st}$ CN in the interval [5.0-5.3] and 3$^{rd}$ CN within [32-34] can in principle contribute to this basin.  Accordingly, three different types of trajectory can be distinguished: 

\begin{enumerate}
\item Successful B4/B3 $\rightarrow$ B1 trajectories, along which extended 5-coordinated regions are present (see Fig.~\ref{Fig3}, 3.9-5.6 ps).
\item Successful B1 $\rightarrow$ B4/B3 trajectories, in which Zn or O are 4-, 5- and 6-coordinated, giving an overall average 1$^{st}$ coordination sphere of five (see Fig.~\ref{Fig2}, 0.7-0.9 ps).

\item Trajectories where the system visits and lingers in the predominantly 5-coordinate intermediate B1-\textit{iH}. The average time spent in this basin is $>$ 200 ps.   
\end{enumerate}

\begin{figure}[t]
 \centering
 \includegraphics[width=9.0 cm]{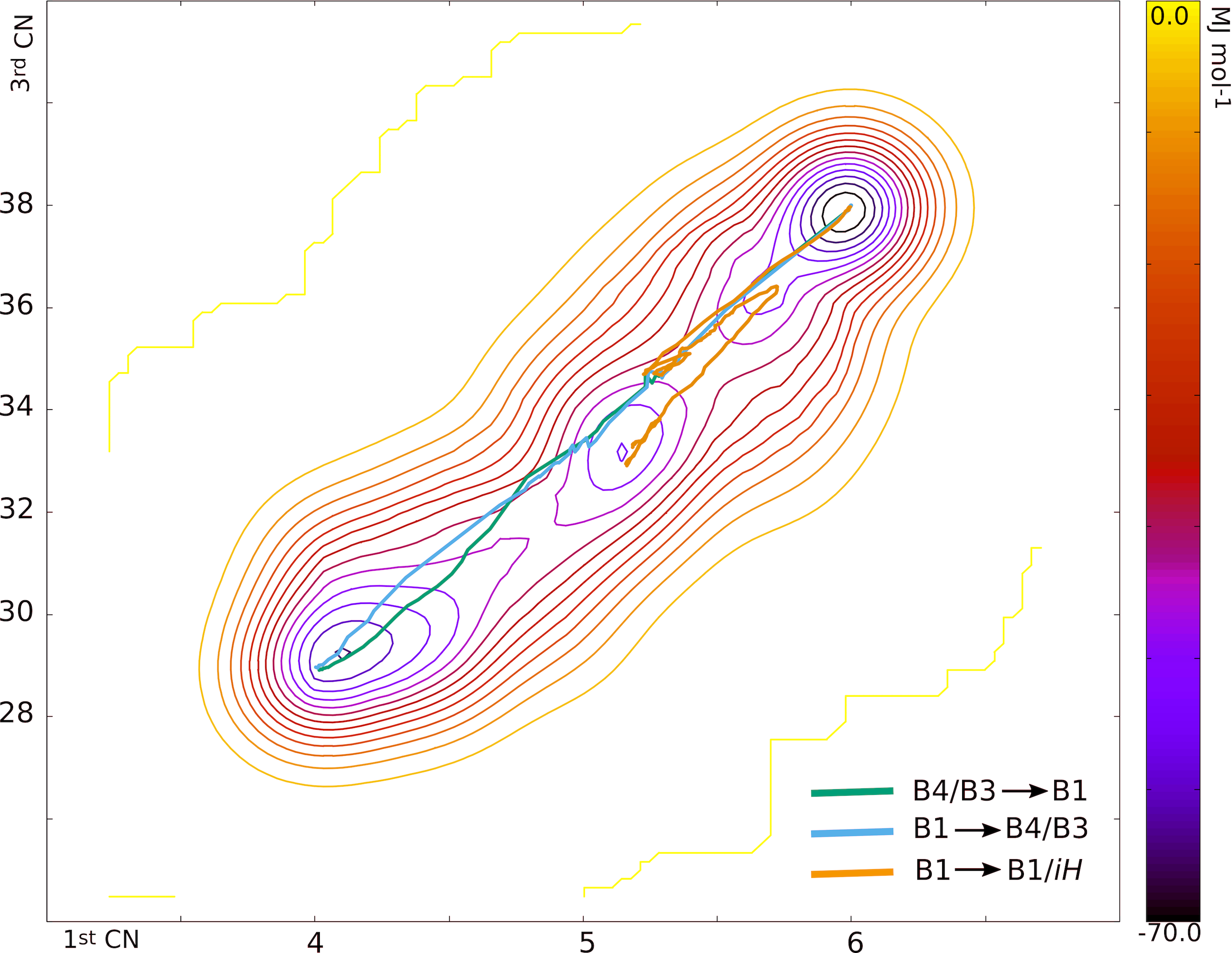}
 \caption{Three representative TPS trajectories, mapped on the free energy surface. Successful B4/B3 $\rightarrow$ B1 trajectory (green),  successful B1 $\rightarrow$ B4/B3 trajectory (light blue) and B1 $\rightarrow$ mixed B1-\textit{iH} (orange). The latter is a long-lived intermediate configuration, which coincides with the intermediate \textit{i}-2 basin.}
 \label{Fig8}
\end{figure}

Representative trajectories of the three types above are mapped on the free energy surface in~Fig.\ref{Fig8}. Successful trajectories tend to keep a higher 3$^{rd}$ coordination number while they approach the large \textit{i-}2 basin. On the other hand, a lower 3$^{rd}$ CN is associated with trajectory entering intermediate basin. Closer to the B1 basin, trajectory are more similar, while the B4/B3 basin can be escaped in different ways. Around TS-B, mechanistic competition (assisted by box size fluctuations~\cite{Boulfelfel:2008fx,Bayarjargal:2014bk}) may result in very different trajectories.

It seems likely, therefore, that the depth of the intermediate basin (and hence the energy barriers associated with it) can not be taken as representing a proper intermediate along a particular type of trajectory. Instead, numerous possible configurations with similar average values of their coordination numbers have all added weight to the free energy at these values of the collective variables, even though not every configuration is realisable in every trajectory.

Unlike TS-A and TS-B, TS-C corresponds to a very `gentle' maximum, with a very shallow gradient either side of it. This again corroborates  what was seen in the path sampling investigations - the system spends a great deal of time in a 4-coordinate state, or a mixture of 4- and 5- coordinate, as a result of a slow, steady process, which maps here onto a gentle free energy gradient. Like its associated intermediate basin \textit{i-}2, the height of TS-C is contributed to by a number of configurations, all of which contain some combination of 4-coordinated ZnO, \textit{iH} intermediate and residual B1. Crossing this maximum implies a significant coordination number (order parameter) change, and corresponds to mixed motif configurations as at 4 ps in Fig.~\ref{Fig2} for the B1 $\rightarrow$ B4/B3 direction, or at 4.7 ps for the B4/B3 $\rightarrow$ B1 (Fig.~\ref{Fig3}).\\

The appearance of mixed B1-\textit{iH} agrees with previous studies on ionic compounds undergoing the same B4/B3$\rightarrow$B1 transition~\cite{Schon:1995ks,Schon:2004bs,Schoen:2015fw}, which may indicate transferability of this mechanistic analysis.  

\subsubsection*{Energy barriers}

\begin{figure}[t]
 \centering
 \includegraphics[width=9.0 cm]{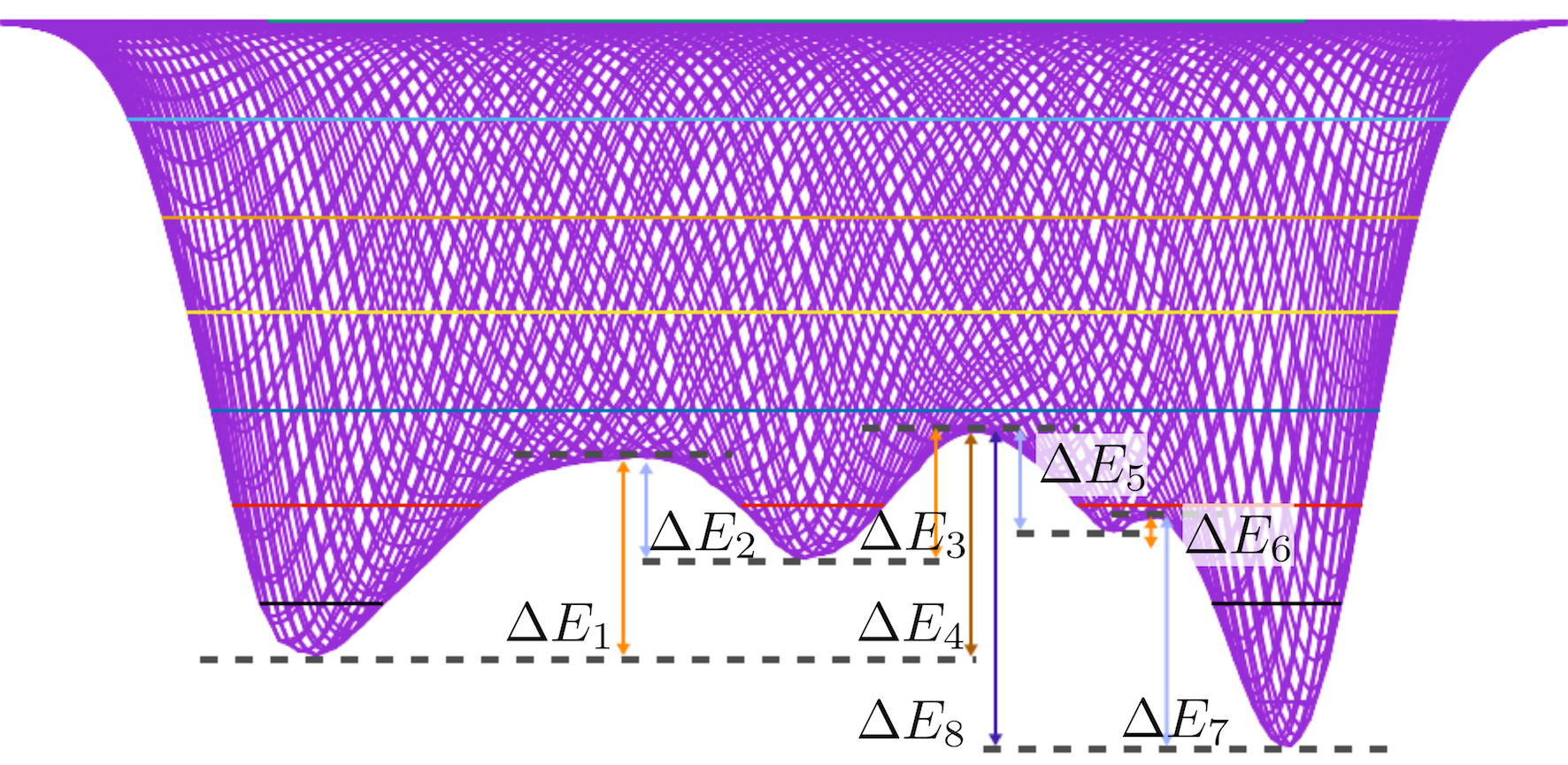}
 \caption{Summary of energy barriers associated with the free energy profile. Activation energies are realised as the largest barrier from the starting basin. Note that these barriers are evaluated as differences from the centres of each of the turning points.}
 \label{Fig7}
\end{figure}

From the converged free energy surface, it is apparent that there are six distinguishable energy barriers associated with the pathway - three for the B4/B3 $\rightarrow$ B1 (Fig.~\ref{Fig7}, E$_{1-3}$), and three for the reverse process (Fig.~\ref{Fig7}, E$_{5-7}$). 

Trajectory variety has shown that the system may proceed along different pathways, therefore crossing different features of the free energy plot. This will lead to different values for the total amount of energy overcome by individual trajectories. The following quantitative analysis of the energy differences assumes measurement to and from the centres of each of the turning points, as described in Table~\ref{table2}, thereby indicating the energy associated with the minimum energy transition pathway that visits all maxima and minima. Every trajectory must surmount an energy barrier at least equivalent to the activation energies presented, as successful trajectories must overcome the highest energy barrier to cross to the other side, irrespective of the path taken.

It is also apparent from inspection that the energy barriers for each stage of the transformation are very different for the forward and reverse trajectories. This may go some way to explaining why there are different pathways associated with the forward and reverse trajectories, and perhaps why there are such notable hysteresis present in such transformations.

The change in free energy is calculated as the difference between the energy values of B1 and B4/B3 basin centres. The intial pressure was chosen based on zero enthalpy difference between B3 and B1, which was dictated by the model used to commence the TPS cycles (see above).

The difference in free energy between the forward (B4/B3 $\rightarrow$ B1) and backward transformation is not zero at the chosen simulation pressure ($p$ = 9.8 GPa). The inclusion of lattice dynamics~\cite{Seko:2005fm} into fist principles calculations was shown to shift the ($\Delta H =0$) "equilibrium" pressure  to a lower value. Consistently with those findings, Fig.~\ref{Fig7} pinpoints the role of entropy, which is positive going from B4/B3 to B1. The disordered nature of the B4/B3 reinforces this effect. 
\\

\begin{table}[h]
\small
  \caption{\ Energy barriers associated with the forward and reverse trajectories. Energy differences are measured to and from the centres of the turning points on the energy surface, corresponding to the (unobserved) minimum energy pathway trajectory}
  \label{table2}
  \begin{tabular*}{0.5\textwidth}{@{\extracolsep{\fill}}lll}
    \hline
    Energy Difference & $kJ mol^{-1}$ & $k_{B}T/ZnO$ \\
    \hline
    \\
    $\Delta E_{1}$ & 20447.0 & 6.8 \\
    $\Delta E_{2}$ & 10483.8 & 3.5 \\
    $\Delta E_{3}$ & 13085.9 & 4.4 \\
    \\
    $\Delta E_{5}$ & 10376.6 & 3.5 \\
    $\Delta E_{6}$ & 1357.0  & 0.5 \\
    $\Delta E_{7}$ & 23726.3 & 7.9 \\
    \\
    $\Delta E_{4}$ & 35069.9 & 11.6 \\
    $\Delta E_{8}$ & 44586.7 & 14.9 \\
    \hline
  \end{tabular*}
\end{table}

\section*{Discussion and Conclusions}

Using transition path sampling methods, the phase transition characterised by a sequence of nucleation and growth events between a mixed wurtzite-zincblende (B4/B3) and rocksalt (B1) has been described. The forward and reverse trajectories were found to be different, each with different competing intermediates and preference for one mechanism over the other for both transformations. 

A typical B1 $\rightarrow$ B4/B3 trajectory was characterised by the formation of an initial seed, followed by gradual transformation to a mixed 4-coordinate product via coexisting motifs. Trajectories could also end up trapped in a long-lived intermediate 5-coordinate basin, corresponding to a mixed rocksalt-5 coordinate hexagonal structure denoted B1-\textit{iH}. The B4/B3 $\rightarrow$ B1 transition entailed a more stepwise process, which involved transition to a 4-coordinate mixture via local motifs of hexagonal \textit{iH} before transforming to the final B1 product. 

A novel method denoted \textit{metashooting} was developed as a combination of TPS and well-tempered metadynamics, using the 1$^{st}$ and 3$^{rd}$ average coordination spheres as the collective variables. TPS ensured rapid variations of the CVs, and a balanced filling of the basins of interest. The \textit{metashooting} procedure is able to decipher the underlying free energy landscape of the transformation, and structural information about the nature of appropriate maxima and minima can be extracted from the resultant plot. In addition, energy barriers and changes in free energy can be ascertained and successful and failed trajectories mapped onto the final free energy plot, in order to gain a true understanding of the energetics and structural changes inherent within each individual trajectory.

\textit{Metashooting} can accommodate additional collective variables, for example higher coordination sequences, in order to further distinguish between the phases. Using the 1$^{st}$ and 3$^{rd}$ coordination spheres was convenient in this case as it was an identical measurement to the order parameter used in the path sampling calculations. In general, the CV set used on the \textit{metashooting} layer can be more detailed then the order parameter utilised in the TPS steps.

It is believed that the \textit{metashooting} procedure, which combines transition path sampling and metadynamics,  could set a novel paradigm for future investigations into condensed matter phase transitions. Such a scheme could easily be transferred to another system, under any level of theory, to ascertain the underlying thermodynamics and kinetics of a temperature or pressure-induced phase transformation.\\

\section*{Acknowledgements}

We would like to thank Davide Branduardi and Michele Parrinello for useful discussions and valuable comments. S.A.J and S.L. would like to thank ARCCA at Cardiff for the generous allocation of computational resources. They also acknowledge support from the UK Research Council for using work in the paper that was undertaken under Project No. EP/M50631X/1. Via our membership of the UK's HPC Materials Chemistry Consortium, which is funded by EPSRC (EP/L000202), this work made use of the facilities of ARCHER, the UK's National High-Performance Computing Service, which is funded by the Office of Science and Technology through EPSRC's High End Computing Programme.\\

\bibliography{references} 
\bibliographystyle{apsrev4-1}
\end{document}